\documentclass[a4paper]{jpconf}
\usepackage{graphicx}
\begin{document}
\title{SU(2) gauge symmetry in gravity phase space}

\author{Francesco Cianfrani.}

\address{Universit\`a di Roma ``Sapienza'', Piazzale Aldo Moro 5, 00185, Roma.}

\ead{francesco.cianfrani@icra.it}

\begin{abstract}
The Hamiltonian formulation of the Holst action in vacuum and in the presence of matter fields is analyzed in a generic local Lorentz frame. It is elucidated how the SU(2) gauge symmetry is inferred by reducing the set of constraints to a first-class one. The consequences of the proposed approach for Loop Quantum Gravity and Spin Foam models are discussed.
\end{abstract}

\section{Introduction}

The reduction of local Lorentz symmetry to SU(2) gauge invariance is the starting point for the whole Loop Quantum Gravity (LQG) framework and it constitutes a key point in order to describe the quantum behavior of the gravitational field via BF theories in Spin-Foam models. In this work we are going to review the procedure by which the Gauss constraints of the SU(2) group arise without fixing the local Lorentz frame in a Hamiltonian 
framework and, then, we will discuss the implications for BF theories.    

\section{Hamiltonian formulation in a generic local Lorentz frame}

Let us consider a space-time manifold endowed with a metric tensor $g_{\mu\nu}$ and fix the local Lorentz frame by 4-bein vectors $e^\alpha_I$ and spin connections $\omega^{IJ}_{\alpha}$. The Holst action reads as follows
\begin{equation}
S=\int \sqrt{-g}\left[e^\alpha_I e^\beta_J R_{\alpha\beta}^{IJ}-\frac{1}{2\gamma}\epsilon^{IJ}_{\phantom{12}KL}e^\alpha_I e^\beta_J R_{\alpha\beta}^{KL}\right]d^4x,
\end{equation}

and conjugate momenta to spin connections are given by
\begin{equation}
{}^\gamma\!\pi_{IJ}^a=\pi_{IJ}^a-\frac{1}{2\gamma}\epsilon^{KL}_{\phantom1\phantom2IJ}\pi_{KL}^a,\qquad \pi_{IJ}^a=2\sqrt{-g}e^t_{[I}e^a_{J]}.
\end{equation}

The Hamiltonian density is a linear combination of the following constraints
\begin{equation}\left\{\begin{array}{c}
H=\pi^a_{IM}\pi^{bM}_{\phantom1J}\left(R^{IJ}_{ab}-\frac{1}{2\gamma}\epsilon^{IJ}_{\phantom1\phantom2KL}R^{KL}_{ab}\right)=0\\
H_a=\pi^b_{IJ}\left(R^{IJ}_{ab}-\frac{1}{2\gamma}\epsilon^{IJ}_{\phantom1\phantom2KL}R^{KL}_{ab}\right)=0\\
G_{IJ}=\partial_a\pi^a_{IJ}-2\omega_{[I\phantom2a}^{\phantom1K}\pi^a_{|K|J]}=0\\
C^{ab}=\epsilon^{IJKL}\pi_{IJ}^{(a}\pi_{KL}^{b)}=0  \\
D^{ab}=\epsilon^{IJKL}\pi^c_{IM}\pi^{(aM}_{\phantom1\phantom2J}D_c\pi^{b)}_{KL}=0
\end{array}\right..
\end{equation}
  
The set of constraints is second-class, which means that some variables are redundant and a non-trivial symplectic structure is inferred on the constraint hypersurfaces.

The standard procedure to deal with such a second-class system is based on fixing the so-called time gauge condition, which means that the boost parameters $\chi_i$ vanish. However, in order to disentangle the fixing of the local Lorentz frame from the solution of second-class constraints, it has been given in [1] a generalized solution of second-class constraint, {\it i.e.}
\begin{equation}
\omega^{\phantom1j}_{i\phantom1a}={}^\pi\!\omega^{\phantom1j}_{i\phantom1a}+\chi_i\omega^{0j}_{\phantom{12}a}+\chi^j
(\omega_{i\phantom1a}^{\phantom10}-{}^\pi\!D_a\chi_i),\qquad\pi^a_{ij}=2\chi_{[i}\pi^a_{0j]}\label{cs1}
\end{equation}

where ${}^\pi\!D_a\chi_i=\partial_a\chi_i-{}^\pi\!\omega^{\phantom1j}_{i\phantom1a}\chi_j$ and $\chi^i=\eta^{ij}\chi_j$. On the hypersurfaces (\ref{cs1}) $\chi_i$ are promoted to configuration variables, such that no gauge fixing of the local Lorentz frame occurs, while second-class constraints are solved. The induced symplectic form is highly nontrivial, but as soon as phase-space is parametrized by densitized triads $E^a_i$ and generalized Ashtekar-Barbero connections $A^i_a$, together with boost paremetrs and conjugate momenta $\pi^i$, the constraints of the Lorentz group are replaced by 
\begin{equation}
G_i=\partial_aE^a_i+\gamma\epsilon_{ij}^{\phantom1k}A^j_aE_k^a=0, \qquad \pi^i=0.\label{gpi}
\end{equation}

Hence, the SU(2) gauge structure arises also when the time-gauge condition is relaxed, while boost parameters are non-dynamical variables and they behave as the lapse function and the shift vector. Henceforth, one can avoid any dependence from $\chi_i$ into the wave functional and perform the quantization as in the standard treatment within the time gauge. The implications of these results on the canonical quantization procedure are that there is no need to perform any gauge fixing of the local Lorentz frame in LQG, while the spectra of geometrical operators is invariant under local boosts. This analysis has been extended to non minimally coupled matter fields [2], Immirzi field [3] and spinor fields [4]. 
 
\section{Rotations and boosts on a quantum level} 

The generators of local Lorentz transformations take the following expression after having solved second-class constraints
\begin{equation}
R_i=G_i+\epsilon_{i\phantom1k}^{\phantom1 j}\chi_j\pi^k, \qquad K_i=(\eta_{ij}+\chi_i\chi_j)\pi^j-\beta\epsilon_i^{\phantom{1}jk}\chi_jG_k.\label{br}
\end{equation}

If one want to analyze the effect of Lorentz transformation on a quantum level, variables $\chi_i$ must be retained into wave functional, such that a generic state on a path $\alpha$ can be written as [5]
\begin{equation}
\psi_\alpha=\otimes_e {}^L\!h_e \otimes_v I_v(\chi^i),\label{st}
\end{equation}

where $e$ and $v$ denote the edges and vertices of $\alpha$, respectively. ${}^L\!h_e$ belongs to the Hilbert space of SU(2) distributional connections, while the Hilbert space associated with boost parameters can be defined as the space of square-integrable functions $I_v(\chi)$ on the hyperbolic space parametrized by $\chi_i$. Symmetric momenta are defined as follows
\begin{equation}
\pi^i=-i\left(\frac{\partial}{\partial\chi_i}+\frac{\chi_i}{2(1-\chi^2)}\right),
\end{equation}

and the scalar product in the full Hilbert space reads as
\begin{eqnarray}
<\psi^1_\alpha,\psi^2_\alpha>=\Pi_{e,v}\int {}^L\!h^{1\dag}_e{}^L\!h^2_e d\mu_{SU(2)} \int \frac{1}{\sqrt{1-\chi^2}} I^{1*}_vI^2_v d^3\chi,
\label{scpr}
\end{eqnarray}

where $d\mu_{SU(2)}$ denotes the Haar measure associated with the SU(2) group. Within this scheme, the wave functionals of LQG can be inferred by imposing Hamiltonian constraints (\ref{gpi}).

In particular, the vanishing of $\pi_i$ gives
\begin{equation}
\pi_i I^{LQG}_v=0\rightarrow I^{LQG}_v(\chi)\propto(1-\chi^2)^{1/4}, \label{pi}
\end{equation}

and once $I^{LQG}$ is inserted into the expression (\ref{scpr}), the scalar product turns out to be LQG one. The condition $G_i=0$ coincides with the Gauss constraint proper of SU(2) gauge theory, thus it can be implemented in the space of distributional connections by inserting invariant intertwiners at vertices. Therefore, it is possible to define the functionals associated with LQG by replacing $I_v(\chi)$ with SU(2) invariant intertwiners. 

The expression (\ref{st}) allow to investigate the effect of boost and rotation generators (\ref{br}). The rotation generators are the sum of the SU(2) generators associated with Ashtekar-Barbero connections and the orbital angular momenta of $\chi^i$. In the case of LQG functionals, $I^{LQG}$ is the representation with vanishing angular momentum, such that the presence of SU(2) invariant intertwiners ensures rotation invariance.
  
As soon as boosts are concerned, let us consider the operator ordering with all momenta on the right (indeed this choice gives a non-symmetric boost generator). Actually, $K_i$ is made of two terms, each one acting at verticies only:
\begin{itemize}
{\item
the first piece contains the conjugate momenta to $\chi_i$ and it acts on the $\chi$-dependent part only as follows (we choose $i=3$) 

\begin{eqnarray*}
(\pi_3+\chi_3\chi_j\pi^j) d_{ln}(\chi^2)Y_{l}^{n}=i(d'_{l n}-l d_{ln})\sqrt{\frac{(l+1)^2-n^2}{(2l+1)(2l+3)}}Y_{l+1}^{n}+i(d'_{ln}+(l+1) d_{l n})\sqrt{\frac{l^2-n^2}{(2l-1)(2l+1)}}Y_{l-1}^{n},
\end{eqnarray*}

where $d'_{ln}=2(1+\chi^2)\partial_\chi^2 d_{l n}-\frac{1}{2}d_{l n}$.}

{\item the second term is made of the SU(2) generators times $\chi_i$ and its action reads ($i=3$)\\\\

$\beta\epsilon_{3}^{\phantom{12}ij}\chi_iG_j |j,m>\otimes d_{ln}Y_{l}^{n}=$\\
$=-\frac{i}{2}\beta\chi \frac{d_{ln}}{\sqrt{2l+1}}\bigg(\sqrt{(j+m)(j-m+1)}\sqrt{(l+n+2)(l+n+1)}|j,m-1>\otimes Y^{n+1}_{l+1}-$\\
$-\sqrt{(j+m)(j-m+1)}\sqrt{(l-n-1)(l-n)}|j,m-1>\otimes Y^{n+1}_{l-1}-$\\
$-\sqrt{(j-m)(j+m+1)}\sqrt{(l-n+2)(l-n+1)}|j,m+1>\otimes Y^{n-1}_{l+1}+$\\
$+\sqrt{(j-m)(j+m+1)}\sqrt{(l+n-1)(l+n)}|j,m+1>\otimes Y^{n-1}_{l-1}\bigg).$}

\end{itemize}
 
As soon as LQG functionals are concerned, the action of the first term vanishes in the adopted operator ordering, while the second term does not provide any contribution because of SU(2) invariance. Therefore, LQG states are invariant under boost transformations.

\section{From BF theory to LQG}

In Spin-Foam models gravity is written as a BF theory, whose action reads
\begin{equation}
S=\frac{1}{2}\int \epsilon^{\alpha\beta\gamma\delta} tr(B^{IJ}_{\alpha\beta}F^{KL}_{\gamma\delta})d^4x,
\end{equation}

with the conditions $B^{IJ}=\left[\delta^{IJ}_{KL}-\frac{1}{2\gamma}\epsilon^{IJ}_{\phantom{12}KL} \right]\ast(e^K\wedge e^L)$. We have seen how such constraints induce on a Hamiltonian level a non trivial symplectic structure, which signals that SU(2) connections $A^i_a$ contain all dynamical information. 
 
In Spin-foam model, holonomies of the full Lorentz group are considered, whose associated irreducible representations of the principal series are made of an infinite tower of SU(2) representations, {\it i.e.}
\begin{equation}
H^{(k,\rho)}=\oplus_{j=k}^{+\infty} H^{(j)},\qquad k\in \frac{\textbf{N}}{2},\qquad \rho\in\textbf{R}.
\end{equation}  

In order to solve consistently second-class constraints one must implement: i) the restriction to the constraint hypersurfaces, ii) the reduction to functionals depending on $A^i_a$ only. This task can be accomplished by taking fully projected spin-network [6], {\it i.e.} 
\begin{equation}
h^{(\lambda,j)}_\alpha=\phantom1^{lim}_{N\rightarrow+\infty}{\it\textsl{P}}\left\{\prod_{n=1}^N\pi^{(j)}h^{(\lambda)}_{e(n)}\pi^{(j)}\right\}, \qquad \alpha=\bigcup_{n=1}^Ne(n),
\end{equation}

where $\pi^{(j)}$ denote the projectors to the representations associated with the Ashtekar-Barbero SU(2) connections, {\it i.e.}
\begin{eqnarray*}
\pi^{(j)}: h^{(k,\rho)}\rightarrow {}^L\!h^j=h^{(j-r,\gamma \frac{j(j+1)}{j-r})},\qquad \gamma=\beta_j=\frac{k\rho}{j(j+1)}. 
\end{eqnarray*}

In fact, the action of momenta gives
\begin{equation}
\pi_{ij}(S) ^L\!h^j_e=0,\qquad \pi_{0i}(S) ^L\!h^j_e=o(S,e) ^L\!h^j_{e_2}\tau_i ^L\!h^j_{e_1}.
\end{equation}

The first relation above signals that the time-gauge condition holds, while the second tells that $\omega^{ij}_a= ^\pi\!\omega^{ij}_a$, thus $^L\!h^j_e$ depend on $A^i_a$ only. Henceforth, the restriction to $ ^L\!h^j_e$ implements properly the features of second-class constraints. 

The condition $\beta_j=\gamma$ fixes the spin number of the selected SU(2) representation inside each Lorentz one.

In a path integral formulation, by adopting fully projected spin-networks, one avoids the nontrivial factor arising from the presence of second class constraints. In fact one finds
\begin{equation}
\int \prod_{ab}\prod_{IJ,KL}d\omega^{IJ}_a d\Pi^b_{KL} \sqrt{det[\chi_\alpha,\chi_\beta]}\Pi_\alpha\delta(\chi_\alpha)f(\omega,\Pi)=\int \prod_{ab}\prod_{ij}dA^i_a dE^b_j f(A,E),
\end{equation}

where $\chi_\alpha=\{C^{ab},D^{ab}\}$. 

Therefore, in order to reduce the phase-space of BF theories, parametrized by holonomies of the Lorentz group, to the one of LQG, it must be addressed inside each Lorentz representation of the principal series the restriction to the SU(2) irreps for which  
\begin{equation}
\rho=\gamma\frac{j(j+1)}{j-r}, \qquad r\in\texttt{N}.
\end{equation}

\end{document}